\documentclass{osa-article}

\journal{oe}
\usepackage{enumitem}
\usepackage{siunitx}

\newcommand{\mb}{\boldsymbol} 

\usepackage{soul,xcolor} \setstcolor{red}
\arrayrulecolor{black}

\newcommand*{\boxedcolor}{blue}
\makeatletter
\renewcommand{\boxed}[1]{\textcolor{\boxedcolor}{%
   \fbox{\normalcolor\m@th$\displaystyle#1$}}}
\makeatother

\newcolumntype{L}{>{\centering\arraybackslash}m{2cm}}

\newcolumntype{K}{>{\centering\arraybackslash}m{8cm}}
\DeclareMathOperator*{\argmin}{arg\,min}

\articletype{Research Article}

\DeclareMathAlphabet{\mathcal}{OMS}{cmsy}{m}{n}
\DeclareMathAlphabet\mathbfcal{OMS}{cmsy}{b}{n}
\begin{document}

\title{Learned reconstructions for practical mask-based lensless imaging}

\author{Kristina Monakhova\authormark{1}, Joshua Yurtsever \authormark{1}, Grace Kuo\authormark{1}, Nick Antipa\authormark{1}, Kyrollos Yanny\authormark{2}, and Laura Waller\authormark{1}}

\address{\authormark{1}Department of Electrical Engineering \& Computer Sciences, University of California, Berkeley, California 94720, USA \\
\authormark{2}Department of Bioengineering, University of California, Berkeley and University of California, San Francisco, Berkeley, CA 94720, USA\\}

\email{\authormark{*}monakhova@berkeley.edu} 



\begin{abstract}
Mask-based lensless imagers are smaller and lighter than traditional lensed cameras. In these imagers, the sensor does not directly record an image of the scene; rather, a computational algorithm reconstructs it. Typically, mask-based lensless imagers use a model-based reconstruction approach that suffers from long compute times and a heavy reliance on both system calibration and heuristically chosen denoisers. In this work, we address these limitations using a bounded-compute, trainable neural network to reconstruct the image. We leverage our knowledge of the physical system by unrolling a traditional model-based optimization algorithm, whose parameters we optimize using experimentally gathered ground-truth data. Optionally, images produced by the unrolled network are then fed into a jointly-trained denoiser. As compared to traditional methods, our architecture achieves better perceptual image quality \textit{and} runs 20$\times$ faster, enabling interactive previewing of the scene.   We explore a spectrum between model-based and deep learning methods, showing the benefits of using an intermediate approach. Finally, we test our network on images taken in the wild with a prototype mask-based camera, demonstrating that our network generalizes to natural images.
\end{abstract}

\section{Introduction}

\begin{figure}[t]
\centering\includegraphics[width=5.25in]{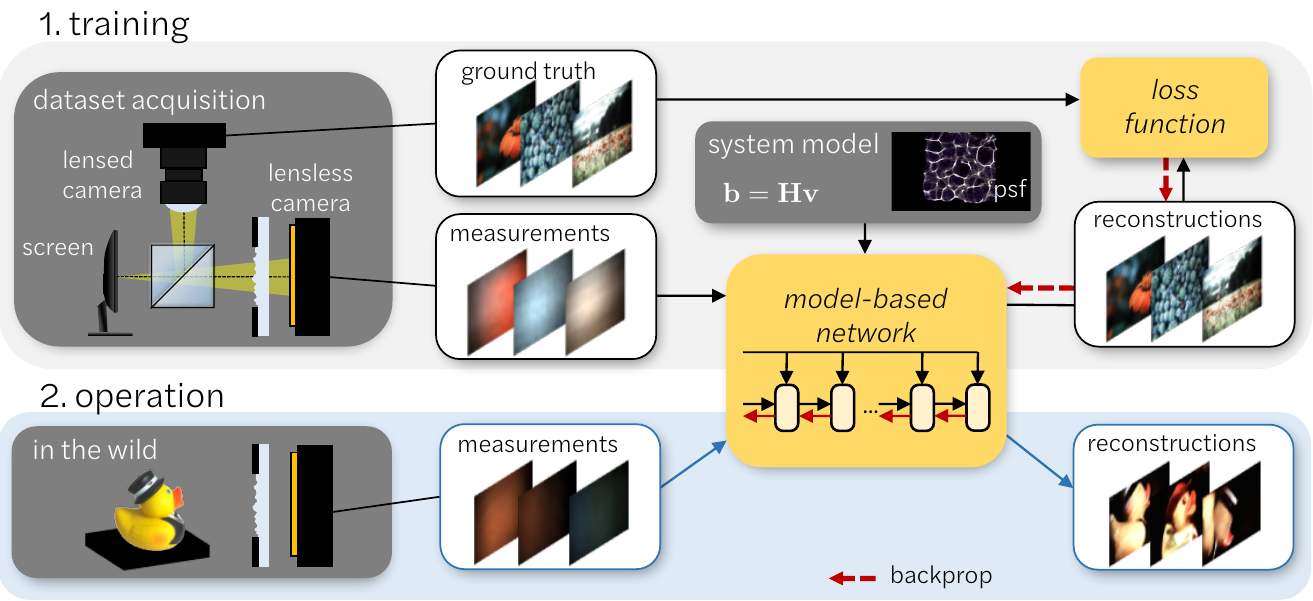}
\caption{Overview of our imaging pipeline. During training, images are displayed on a computer screen and captured simultaneously with both a lensed and a lensless camera to form training pairs, with the lensed images serving as ground truth labels.  The lensless measurements are fed into a model-based network which incorporates knowledge about the physics of the imager. The output of the network is compared with the labels using a loss function and the network parameters are updated through backpropagation.  During operation, the lensless imager takes measurements and the trained model-based network is used to reconstruct the images, providing a large speedup in reconstruction time and an improvement in image quality. }
\label{fig:overview}
\end{figure}

Mask-based lensless imagers (lensless imagers) are a class of computational cameras in which the lens is replaced with a phase or amplitude mask placed a short distance in front of the sensor (Fig.~\ref{fig:overview}). Unlike conventional (lensed) cameras, which directly record an image, lensless cameras map each point in the scene to many sensor pixels, \textit{indirectly} encoding scene information into the sensor measurement. A reconstruction algorithm is then used to recover the final image. This architecture enables small, cheap, and light-weight designs which can be used for portable or \textit{in vivo} imaging~\cite{asif2015flatcam, TOMB0_tanida2001thin, antipa2018diffusercam, kuo20183d, yanny2019miniature, liu2019single}.  Additionally, the inherent multiplexing of lensless cameras can make them amenable to compressive measurement of higher-dimensional signals, such as 3D volumetric~\cite{antipa2018diffusercam, TOMBO_3d_Horisaki2007} or video~\cite{antipa2019video}, from a single 2D measurement. Lensless cameras have been used for 3D fluorescence microscopy~\cite{flatscope_rice, kuo20183d}, thermal imaging~\cite{gill2017thermal}, and refocusable photography~\cite{Tajima_iccp2017_lenslessLF}.  

Image reconstruction methods for lensless cameras fall into two general categories: single-step and iterative reconstructions.  Single-step reconstructions can be fast, but often require custom fabricated masks that must be carefully aligned to the sensor~\cite{asif2015flatcam, Tajima_iccp2017_lenslessLF, gill2017thermal, TOMB0_tanida2001thin}.  In addition, it is difficult to incorporate priors and leverage compressed sensing in single-step reconstructions. Iterative reconstructions are much slower, but do not impose stringent restrictions on the mask itself, generally produce better results, and allow priors to be used~\cite{kuo2017diffusercam, RAMBUS}. However, due to imperfect system modeling, these methods may still give significant reconstruction artifacts.  Additionally, the high complexity of the computation precludes interactive previewing of the scene and requires expensive, bulky compute hardware.  In this work, we focus on iterative methods, improving both the image quality and speed with a new reconstruction framework that incorporates the advantages of both deep learning and physical models, making lensless cameras more practical for everyday imaging.
 
The classical approach to image recovery is to use convex optimization to iteratively minimize a loss function~\cite{beck2009fast, boyd2011distributed} consisting of a data-fidelity term and an optional hand-picked regularization term. The data-fidelity term enforces that the recovered image, with the known imaging model applied to it, matches the measurement. The regularization term enforces prior knowledge of image statistics (e.g. non-negative, sparse gradients) and serves to regularize ill-conditioned problems.   Iterative approaches are interpretable, but are sensitive to reconstruction artifacts due to model mismatch, calibration errors, hand-tuned parameters, and hand-picked regularizers which are not necessarily representative of the data.  Each of these contributes to reconstruction artifacts and degrades image quality.  Furthermore, these methods can take hundreds to thousands of iterations to converge, which is often too slow for real-time imaging.  

Recently, deep learning-based methods for image reconstruction have risen in popularity.  In deep methods, a convolutional neural network (CNN) is used for image reconstruction~\cite{li2018imaging, li2018deep, nguyen2018deep}. Networks have hundreds of thousands of parameters which are updated using large datasets of image pairs. These networks are able to learn complex scene statistics, but do not incorporate any prior knowledge about the image formation process. Compared to iterative methods, deep learning-based methods are hard to interpret, do not have convergence guarantees, and have no structured way to incorporate knowledge of the imaging system physics. 

Unrolled optimization represents a middle-ground between classic and deep methods.  In unrolled optimization, a fixed number of iterations from a classic algorithm is interpreted as a deep network, with each iteration serving as a layer in the network.  In each layer, if the parameters of the algorithm are differentiable with respect to the output, they can be optimized for a given loss function through backpropagation.  In this framework, the sparsifying filters, hyper-parameters, or shrinkage function can be learned from the training examples~\cite{gregor2010learning, schmidt2014shrinkage}.  Unrolled optimization has shown promising results for image denoising~\cite{diamond2017dirty, diamond2017unrolled}, sparse coding~\cite{gregor2010learning}, and MRI reconstructions~\cite{sun2016deep}.

\begin{figure}[t]
\centering\includegraphics[width=5.25in]{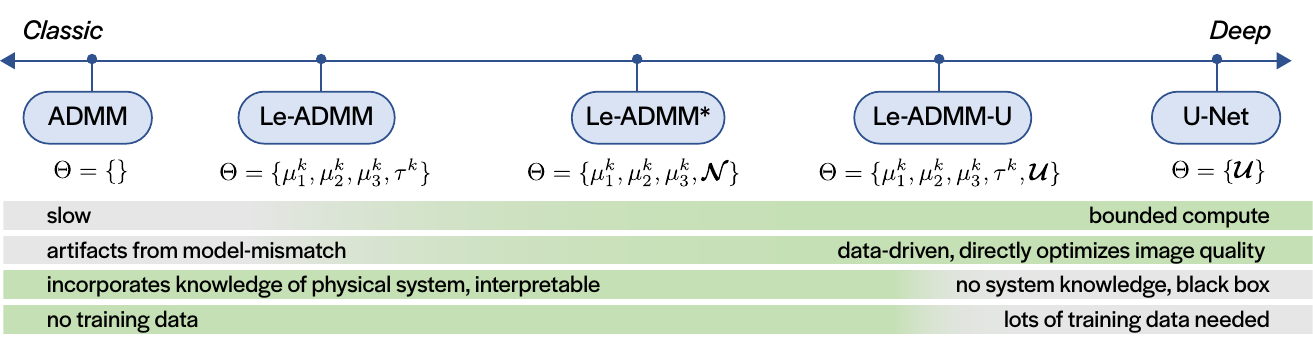}
\caption{Networks on a scale from classic to deep.  We will present several networks specifically designed for lensless imaging (Le-ADMM, Le-ADMM*, and Le-ADMM-U).  We compare these to classic approaches, which have no learnable parameters, and to purely deep methods which do not include any knowledge of the imaging model. We will show the utility of using an algorithm in this middle range compared to a purely classic or deep method.  $\Theta$ summarizes the parameters that are learned for each network as discussed in Section \ref{sec:networks}.}
\label{fig:classic2deep}
\end{figure}

Here, we unroll the iterative \textit{alternating direction method of multipliers} (ADMM) algorithm with a variable splitting specific for lensless imaging~\cite{boyd2011distributed, antipa2018diffusercam}.  This allows us to incorporate knowledge of the image formation process into the neural network as well as learn the network parameters based on the data.  To train our network, we experimentally capture a large dataset of lensed and lensless images (Fig.~\ref{fig:overview}).  We train our network on a perceptual similarity metric in order to produce images that are visually similar to those from our ground truth lensed camera.  We present several variations of networks along the spectrum between classic methods and deep methods, by varying the amount of trainable parameters (Fig.~\ref{fig:classic2deep}).  Specifically, we introduce three architectures, Le-ADMM, Le-ADMM*, and Le-ADMM-U, each with increasing numbers of trainable parameters, explained in detail in Sec.~\ref{sec:networks}. All of our networks have a bounded compute that can be adjusted according to the application.  The networks trade-off data fidelity and image perceptual quality, producing more visually appealing images at the price of decreased data-fidelity.

We test our network using DiffuserCam~\cite{kuo2017diffusercam} as our prototypical lensless camera, built with off-the-shelf components and a low-end camera sensor. Although our network is trained using images from a computer monitor, we demonstrate the generalization of our network to measurements of natural objects taken in the wild.  We believe that this exploratory work shows the promise of using unrolled neural networks for lensless imaging, and our results suggest the utility of combining knowledge of the physics together with deep learning for the best performance.   

Our contributions include:
\begin{enumerate}[noitemsep,nolistsep]
    \item A bounded-time trainable network architecture that incorporates knowledge of the physical model for lensless imaging.
    \item An experimental dataset of 25,000 aligned lensed and lensless image pairs taken using a beamsplitter and computer screen.
    \item A demonstration of 20$\times$ speedup and 3$\times$ improvement in perceptual similarity for lensless imaging reconstructions on an experimental system.
    \item Generalization of the network to images taken in the wild on a prototype lensless camera.
\end{enumerate}

\section{Lensless imaging forward model}

First we describe our lensless imaging forward model for DiffuserCam. Based on this, we formulate our traditional model-based reconstruction (Sec.~\ref{sec:inverse}), before moving on to modifications that span the spectrum from model-based to deep learning-based algorithms (Fig.~\ref{fig:classic2deep}) in Sec.~\ref{sec:networks}.

DiffuserCam~\cite{antipa2018diffusercam,kuo2017diffusercam} is a compact, easy-to-build imaging system that consists only of a diffuser (a transparent phase mask with pseudo-random slowly varying thickness) placed a few millimeters in front of a standard image sensor (see Fig.~\ref{fig:overview}). Light from a point source in the scene is refracted by the diffuser to create a high-contrast caustic pattern on the sensor, which is the point spread function (PSF) of the system (Fig.~\ref{fig:overview}). Since the diffuser is thin, the PSF can be modeled as shift-invariant: a lateral shift of the point source in the scene causes a translation of the PSF in the opposite direction. We model the scene as a collection of point sources with varying color and intensity. Assuming all points are incoherent with each other, the sensor measurement, $
\mathbf{b}$, can be described as:

\begin{align}
\label{eqn:forward}
\begin{split}
    \mathbf b(x,y)  &=  \text{crop} \lbrack \mathbf h(x,y) * \mathbf x(x,y) \rbrack \\
   & = \mathbf C \mathbf H \mathbf x ,
\end{split}
\end{align}

\noindent where $\mathbf h$ is the system PSF, $\mathbf x$ represents the scene, and $(x,y)$ are the sensor coordinates. Here, $*$ denotes 2D discrete linear convolution, which returns an array that is larger than both the scene and the PSF. Therefore, a crop operation restricts the output to the physical sensor size. This relation is represented compactly in matrix-vector notation with $\text{crop}$ denoted as $\mathbf C$ and convolution with the PSF denoted as $\mathbf H$.  Equation~\eqref{eqn:forward} is computed separately for each color channel.

Our goal is to recover the scene, $\mathbf x$, from the measurement $\mathbf b$. We assume the PSF is known, as it can easily be measured experimentally with an LED point source~\cite{antipa2018diffusercam}. Traditional model-based methods for recovering $\mathbf x$ solve a regularized optimization problem of the following form:

\begin{equation}
\label{eqn:backward}
    \hat{\mathbf x} = \argmin_{\mathbf x\geq 0} \frac{1}{2} \| \mathbf b -  \mathbf {C H} \mathbf x \|_2^2 + \tau \|  {\Psi} \mathbf x\|_1,
\end{equation}

\noindent where $\Psi$ is a sparsifying transform, such as finite differences for total variation (TV) denoising, and $\tau$ is a tuning parameter that adjusts the sparsity level.  

\section{Model-based inverse algorithm} \label{sec:inverse}

The traditional model-based inverse solver relies on the known physics of the forward model to solve Eq.~\eqref{eqn:backward}, minimizing the difference between the actual and predicted measurements, while satisfying any additional constraints. This problem can be solved efficiently by ADMM~\cite{boyd2011distributed} with a variable splitting that leverages the structure of the problem~\cite{antipa2018diffusercam}.  In ADMM, the problem is reformulated as:

\begin{align}
\label{eqn:backward_admm}
\begin{split} 
\hat{\mathbf x} = \argmin_{w\geq 0, u, v} \frac{1}{2} \| \mathbf b -  \mathbf C v \|_2^2 + \tau \| u\|_1,\\
s.t. \; v=  \mathbf H  \mathbf x, u = \Psi  \mathbf x, w =  \mathbf x.
\end{split}
\end{align}

\noindent This variable splitting allows closed-form updates for each step, as derived in~\cite{antipa2018diffusercam}.  The update equations in each iteration become: 

\begin{align*}
u^{k+1}& \leftarrow \mathcal{T}_{\tau/\mu_2}(\mb \Psi \mathbf{x}^k + \alpha_2^k/\mu_2) &\text{sparsifying soft-threshold}\\  
v ^{k+1}& \leftarrow (\mathbf C^{\mathbf T} \mathbf C + \mu_1I)^{-1}(\alpha_1^k + \mu_1 \mathbf H \mathbf x^k + \mathbf C^{\mathbf T} \mathbf b) &\text{least-squares update}\\ 
w^{k+1} &\leftarrow \text{max}(\alpha_3^k/\mu_3 + \mathbf x^k, 0)   &\text{enforce non-negativity} \\
\mathbf x^{k+1} &\leftarrow  (\mu_1 \mathbf H^{\mathbf T} \mathbf H + \mu_2 \mb \Psi^{\mathbf T} \mb\Psi + \mu_3I)^{-1} r^k  &\text{least-squares update}\\
\alpha_1^{k+1} &\leftarrow \alpha_1^{k} + \mu_1(\mathbf H \mathbf x^{k+1} - v^{k+1})  &\text{dual for }v\\ 
\alpha_2^{k+1} &\leftarrow \alpha_2^{k} + \mu_2(\mb\Psi \mathbf x^{k+1} - u^{k+1}) &\text{dual for }u\\ 
\alpha_3^{k+1} &\leftarrow \alpha_3^{k} + \mu_3(\mathbf x^{k+1} - w^{k+1}) &\text{dual for }w\\ 
\text{where } r^k &= ((\mu_3w^{k+1}-\alpha_3^k) + \mb \Psi^{\mathbf T}(\mu_2 u^{k+1} - \alpha_2^k) + \mathbf H^{\mathbf T}(\mu_1 v^{k+1} - \alpha_1^k)).
\end{align*}

\noindent Here, $\alpha_1$, $\alpha_2$, and $\alpha_3$ are the Lagrange multipliers, or dual variables, respectively associated with $u$, $v$, and $w$, and  $\mu_1$, $\mu_2$, and $\mu_3$ are scalar penalty parameters. $\mathcal{T}_{\tau/\mu_2}$ denotes vectorial soft-thresholding with parameter $\tau/\mu_2$.

This traditional method is based on the physical model of the imaging system (Eq.~\eqref{eqn:forward}) and requires no additional calibration data beyond the PSF. However, it depends heavily on hand-chosen values, such as the sparsifying transform $\Psi$ and its associated parameter, $\tau$.  The optimization parameters, $\mu_1$, $\mu_2$, and $\mu_3$, are either hand-tuned or auto-tuned based on the primal and dual residuals at each iteration ~\cite{boyd2011distributed}.  The method performs well under correctly chosen sparsifying transforms and with the proper hand-tuned parameters.  However, in practice ADMM takes hundreds of iterations to converge and produces images with reconstruction artifacts.  In the next section, we will outline how we unroll ADMM into a neural network in order to learn the hyper-parameters from the data and seamlessly interface with existing deep learning pipelines.  


\section{Learned reconstruction networks} \label{sec:networks}

\begin{figure}[t]
\centering\includegraphics[width=5.25in]{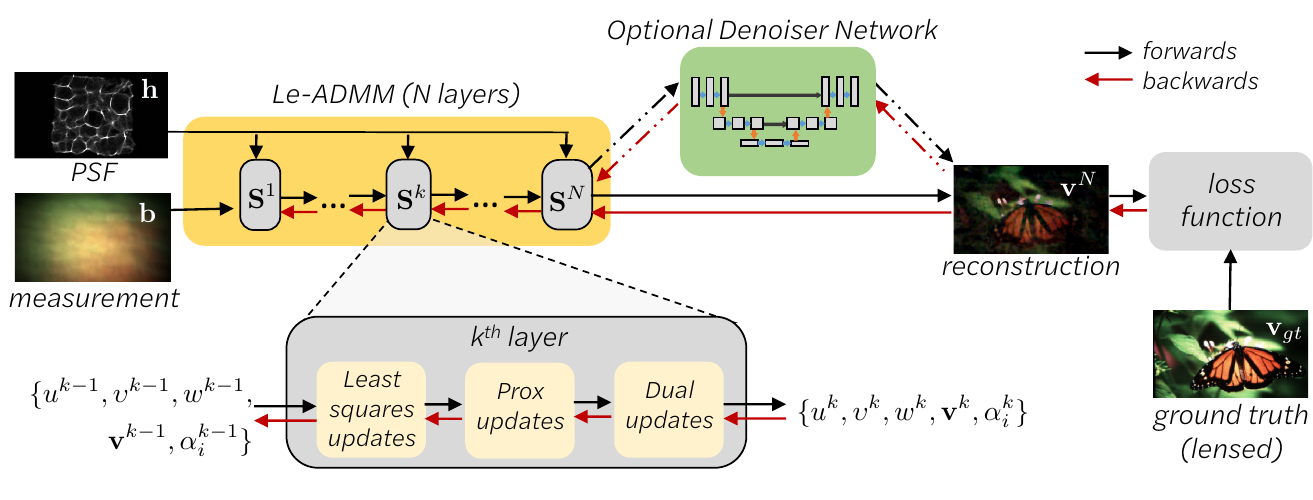}
\caption{Model-based Network Architecture. The input measurement and the calibration PSF are first fed into N layers of unrolled Le-ADMM. At each layer, the updates corresponding to $\mathbf S^{k+1}$ in Eq. \eqref{eqn:leadmm} are applied.  The output of this can be fed into an optional denoiser network.  The network parameters are updated based on a loss function comparing the output image to the lensed image.  Red arrows represent backpropagation through the network parameters.}
\label{fig:network}
\end{figure}

Next, we present several variations of neural networks that jointly incorporate known physical models and deep learning principles.  Each network is based on unrolling the iterative ADMM algorithm, such that each iteration comprises a layer of the network, with the tunable parameters learned from the training data. Thus, the physical model is inherently built into the network architecture, making it more efficient. 

We present three variations of networks, each having a different number of learned parameters.  Learned ADMM (Le-ADMM) has trainable tuning and hyper-parameters.  Le-ADMM* extends Le-ADMM by adding a trainable CNN instead of a hand-tuned sparsifying transform.  Finally, Le-ADMM-U adds a trainable deep denoiser based on a CNN as the last layer of the Le-ADMM network, learning both the hyper-parameters of Le-ADMM as well as the denoiser.  Figure~\ref{fig:classic2deep} summarizes these methods and where they fall on a scale from classic to deep, and the following sections describe them in detail.  Each method has progressively more trainable parameters, and therefore needs a larger training dataset.  All networks use 5 iterations of unrolled ADMM, in order to target a 20$\times$ speed improvement, which would speed up each reconstruction from \SI{1.5}{\second} to \SI{75}{\milli\second}, giving a practical speed for real world imaging. 

\subsection{Learned AMMM (Le-ADMM)}
In the simplest of our unrolled networks, Le-ADMM (learned ADMM), we model each $k^{\textnormal{th}}$ iteration of ADMM as a layer in a neural network, outlined in Fig.~\ref{fig:network}. In Le-ADMM, the optional denoiser step depicted in Fig.~\ref{fig:network} is omitted.  We denote the collection of update equations at the $k^{\textnormal{th}}$ step of ADMM as $\mathbf S^k$. These update Eqs. are given by: 

\begin{align}
\begin{split}
\mathbf S^{k+1} \leftarrow 
\begin{cases}
u^{k+1} \leftarrow \mathcal{T}_{\boxed{\tau^k}}( \mb\Psi(\mathbf{x}^k) + \alpha_2^k/\boxed{\mu_2^k}) & \text{sparsifying soft-thresholding}\\  
v ^{k+1} \leftarrow (\mathbf C^{\mathbf T} \mathbf C + \mu_1I)^{-1}(\alpha_1^k + \boxed{\mu_1^k} \mathbf H \mathbf x^k + \mathbf C^{\mathbf T} \mathbf b) & \text{least-squares update}\\ 
w^{k+1} \leftarrow \text{max}(\alpha_3^k/ \boxed{\mu_3^k} + \mb x^k, 0)  & \text{enforce non-negativity} \\
\mathbf x^{k+1} \leftarrow (\boxed{\mu_1^k} \mathbf H^{\mathbf T} \mathbf H + \boxed{\mu_2^k} \mb\Psi^{\mathbf T} \mb\Psi + \boxed{\mu_3^k}I)^{-1} r^k  & \text{least-squares update} \\
\alpha_1^{k+1} \leftarrow \alpha_1^{k} + \boxed{\mu_1^k}(\mathbf H \mathbf x^{k+1} - v^{k+1})  & \text{dual for v}\\ 
\alpha_2^{k+1} \leftarrow \alpha_2^{k} + \boxed{\mu_2^k}( \mb\Psi(\mathbf x^{k+1}) - u^{k+1}) & \text{dual for u} \\ 
\alpha_3^{k+1} \leftarrow \alpha_3^{k} + \boxed{\mu_3^k}(\mathbf x^{k+1} - w^{k+1}) & \text{dual for w} \\ 
\end{cases}   \\
\text{where }  r^k = ((\boxed{\mu_3^k}w^{k+1}-\alpha_3^k) +  \mb\Psi^{\mathbf T}(\boxed{\mu_2^k} u^{k+1} - \alpha_2^k) + \mathbf H^{\mathbf T}(\boxed{\mu_1^k} v^{k+1} - \alpha_1^k)).
\label{eqn:leadmm}
\end{split}
\end{align}

\noindent The trainable parameters are outlined in blue and can be summarized by $\Theta= \{\mu_1^k, \mu_2^k, \mu_3^k, \tau^k \}$, where $k$ represents the iteration number. For 5 unrolled layers, we have a total of 20 learned parameters. After a fixed number of ADMM iterations the reconstruction is compared to the ground truth (lensed) image using the loss function described in Section~\ref{sec:losses}.  The trainable parameters are updated using backpropagation to minimize this loss across multiple training examples. Le-ADMM can be interpreted as a data-tuned ADMM where the parameters that are typically hand-tuned or auto-tuned are now updated based on the data in order to minimize a data-driven loss function.

\subsection{Le-ADMM*, with learned regularizer}
Le-ADMM* has the same overall structure as Le-ADMM, but also includes a learnable regularizer based on a CNN.   The new update steps are summarized below:

\begin{align}
\begin{split}
\mathbf S^{k+1} \leftarrow 
\begin{cases}
u^{k+1} \leftarrow \boxed{\mathbfcal{N}}(\mathbf x^k) & \text{network regularizer}\\  
v ^{k+1} \leftarrow (\mathbf C^{\mathbf T} \mathbf C + \mu_1I)^{-1}(\alpha_1^k + \boxed{\mu_1^k} \mathbf H \mathbf x^k + \mathbf C^{\mathbf T} \mathbf b) & \text{least-squares update}\\ 
w^{k+1} \leftarrow \text{max}(\alpha_3^k/ \boxed{\mu_3^k} + \mb x^k, 0)  & \text{enforce non-negativity} \\
\mathbf x^{k+1} \leftarrow (\boxed{\mu_1^k} \mathbf H^{\mathbf T} \mathbf H + \boxed{\mu_2^k} I + \boxed{\mu_3^k}I)^{-1} r^k  & \text{least-squares update} \\
\alpha_1^{k+1} \leftarrow \alpha_1^{k} + \boxed{\mu_1^k}(\mathbf H \mathbf x^{k+1} - v^{k+1})  & \text{dual for v}\\ 
\alpha_3^{k+1} \leftarrow \alpha_3^{k} + \boxed{\mu_3^k}(\mathbf x^{k+1} - w^{k+1}) & \text{dual for w} \\ 
\end{cases}   \\
\text{where }  r^k = ((\boxed{\mu_3^k}w^{k+1}-\alpha_3^k) + \boxed{\mu_2^k} u^{k+1} + \mathbf H^{\mathbf T}(\boxed{\mu_1^k} v^{k+1} - \alpha_1^k)).
\label{eqn:leadmms}
\end{split}
\end{align}

\noindent $\mathbfcal{N}$ represents a learnable network, applied at each ADMM iteration. For this learnable transform, we use a small U-Net based on~\cite{ronneberger2015u} consisting of a single encoding and decoding step (complete architecture is available in the Appendix). Because $\mathbfcal{N}$ does not represent the solution to a well defined convex problem, we drop $\alpha_2$, the dual variable associated with $u$. The learnable parameters for Le-ADMM* are thus given by $\Theta= \{\mu_1^k, \mu_2^k, \mu_3^k,  \mathbfcal{N} \}$, which is a total of 32,135 learned parameters. The added learned parameters allow Le-ADMM* to learn a better prior on the data and account for model-mismatch errors in the forward model, at the price of requiring additional training data.

\subsection{Le-ADMM-U}
Our third variation of unrolled networks is Le-ADMM followed by a learned denoiser, as shown in Fig.~\ref{fig:network}.  Here, a U-Net is used as the denoiser~\cite{ronneberger2015u}.  This method has the most learnable parameters, having a total of 10,605,927 learned parameters, all but 20 of which are from the U-Net.  The parameters of Le-ADMM-U, given by $\Theta= \{\mu_1^k, \mu_2^k, \mu_3^k, \tau^k,  \mathbfcal{U} \}$, are jointly updated throughout training.  The Le-ADMM portion of the network performs the bulk of the deconvolution and includes knowledge of the forward model, while the U-Net denoises the final image, is able to correct model mismatch errors, and makes the images look more visually appealing. Our denoiser network architecture is described in the Appendix.

\subsection{U-Net}
For completeness, we also compare to a purely deep method with no knowledge of the system physics in the reconstruction. For this, we directly use the U-Net architecture from~\cite{ronneberger2015u}, resulting in 10,605,907 learned parameters. We summarize this network architecture in the Appendix. 

\subsection{Loss functions} \label{sec:losses}
The loss function must be carefully selected because it dictates the parameter updates throughout the training process.  In classic methods, ground truth is unavailable, so the loss is a function of the consistency of the final image, $\hat{\mathbf x}$ with the measurement model and any image priors. With the inclusion of ground truth training data pairs, we now have access to another class of loss functions that directly compare a given reconstructed image to its associated ground truth image, $\mathbf x_{gt}$.  One common loss is the mean-squared error (MSE) loss with respect to the ground truth, $\|\mathbf x_{gt} -  \hat{\mathbf x}\|^2_2$.  However, MSE favors low frequencies and generally results in learned reconstructions that are blurry and lack detail~\cite{zhang2018unreasonable}. Here, we will use the Learned Perceptual Image Patch Similarity metric (LPIPS) that uses deep features and aims to quantify a perceptual distance between two images, as introduced in~\cite{zhang2018unreasonable}.  During training, we use a combination of both MSE and LPIPS, as outlined in Section~\ref{Implementation}. These loss functions are summarized in Table~\ref{tab:loss}.


\begin{table}[h]
    \centering
    \caption{Loss functions. We use a combination of MSE and LPIPS during training of the learned methods. In the classical methods, there is no ground truth data, $\mathbf{x}_{gt}$, so data fidelity is used with total variation for regularization.}
    \begin{tabular}{L L K} 
         \textbf{Data Fidelity} & $\frac{1}{2}\|\mathbf b - \mathbf C \mathbf H \hat{\mathbf x}\|_2^2$ & Consistency of the measurement with our knowledge of the imaging system  \\ \hline
         \textbf{MSE}&  $\|\mathbf x_{gt} -  \hat{\mathbf x}\|^2_2$ & Pixel-wise difference between reconstruction and ground truth  \\ \hline
         \textbf{LPIPS} & LPIPS($\mathbf x_{gt}, \hat{\mathbf x} $)& Perceptual distance between reconstruction and ground truth~\cite{zhang2018unreasonable} \\ 
    \end{tabular}
    \label{tab:loss}
\end{table}

\section{Implementation}\label{Implementation}

For training, we simultaneously collect a set of lensless and ground truth image pairs using an experimental setup consisting of a lensed camera, a DiffuserCam, a beamsplitter, and computer monitor (Fig.~\ref{fig:overview}).  The cameras and computer monitor are simultaneously triggered, which allows us to display and capture all the training pairs in the dataset overnight.  

Our DiffuserCam prototype consists of an off-the-shelf diffuser (Luminit 0.5$^{\circ}$) with a laser-cut paper aperture placed approximately 9 mm from a CMOS sensor.  The lensed camera is focused at the plane of the computer screen, approximately 10 cm away. We capture a calibration PSF (see Fig.~\ref{fig:overview}) using an LED point source placed at the distance of the computer screen, which sets the focal plane of the DiffuserCam. For both DiffuserCam and the ground truth camera, we use Basler Dart (daA1920-30uc) sensors.  We use a \SI{6}{\milli\metre} S-mount lens for the ground truth camera and calibrate the lens distortion using OpenCV's undistort camera calibration procedure \cite{opencv_library}.  To achieve pixel-wise alignment between the image pairs, we first optically align the two cameras, then further calibrate by displaying a series of points on the computer monitor that span the field-of-view. We reconstruct these point images and compute the homography transform needed to co-align both cameras' coordinate systems. This transform is applied to all subsequent images.  

Our dataset consists of 25,000 images from the MirFlickr dataset~\cite{huiskes08}.  The raw data from each sensor is 1920$\times$1080 pixels, but is down-sampled by a factor of 4 in each direction, to 480$\times$270. This is necessary due to moire fringes from the screen which degrade our lensed image quality. We split the dataset up into 24,000 training images and 1,000 test images.   Our networks are implemented in PyTorch and trained on a Titan X GPU, using an ADAM optimizer throughout training~\cite{kingma2014adam}.  We find that using a combined loss based on MSE and LPIPS works best in practice.  We weight MSE more heavily during earlier epochs and weight LPIPS more heavily during later epochs for further refinement. Source code is available at \cite{lenslessLearningCode}.  When displaying the final images, we crop to 380$\times$210 pixels to avoid displaying areas beyond the borders of the computer monitor.

\section{Results}

After training, we compare the performance of our unrolled networks against both classic ADMM and the fully deep U-Net.  Since the number of iterations of ADMM affects both speed and quality of the result, we compare against both ADMM run until convergence (100 iterations) as well as ADMM bounded to 5 iterations.  Bounded ADMM takes a similar time to run as our unrolled networks and converged ADMM sets a baseline for the best performance classic algorithms can achieve.  On the deep side, we compare against a U-Net which is trained using our raw DiffuserCam measurements and ground truth labels.  

The reconstruction results of images in our test set (taken by the monitor setup, but not used during training) show that our fastest learned networks are able to produce similar or better images than converged ADMM in the same amount of time as bounded ADMM (5 iterations), a 20$\times$ speedup while achieving comparable or better image quality.  Furthermore, we show reconstructions of natural images in the wild (not from a computer monitor), demonstrating that our networks are able to generalize to 3D objects with variable lighting conditions.

\begin{figure}[t]
\centering\includegraphics[width=5.25in]{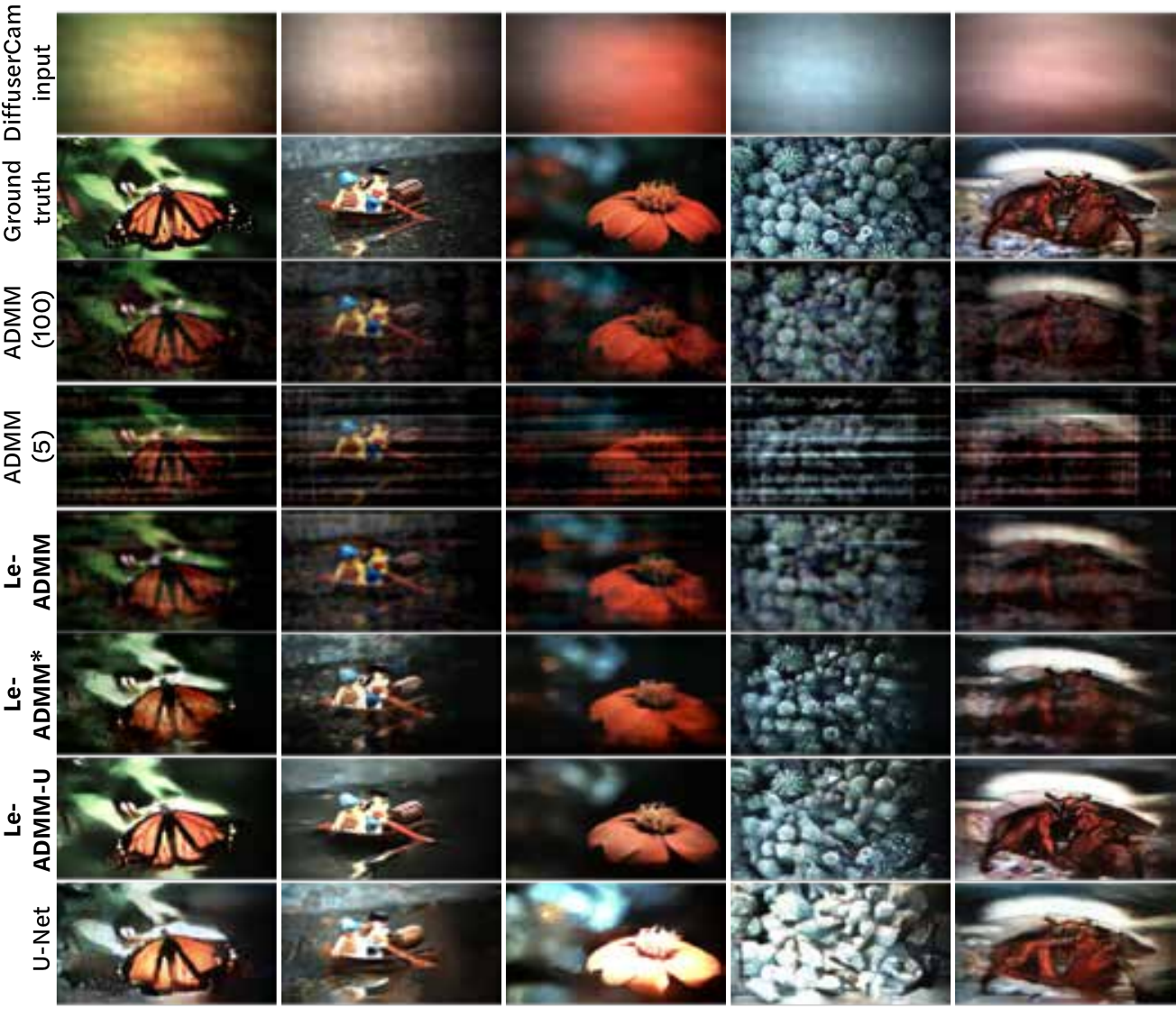}
\caption{Test set results, with the raw DiffuserCam measurement (contrast stretched) and the ground truth images from the lensed camera for reference.  Le-ADMM (\SI{71}{\milli\second}) has similar image quality to converged ADMM (\SI{1.5}{\second}) and better image quality than bounded ADMM (\SI{71}{\milli\second}).  Le-ADMM* and Le-ADMM-U have noticeably better visual image quality.  The U-Net by itself is unable to reconstruct the appropriate colors and lacks detail. }
\label{fig:screenresults}
\end{figure}

\begin{table}[h]
    \centering
    \caption{Network performance. We summarize the average data fidelity, MSE, and LPIPS metrics for each network on the test set (1,000 images). Le-ADMM and Le-ADMM-U are both 20$\times$ faster than converged ADMM with comparable or better performance in terms of MSE and LPIPS.  Le-ADMM-U has the best performance in terms of MSE and LPIPS, outperforming the U-Net which has no knowledge of the system physics.}
    \vspace{2mm}
    \begin{tabular}{c c c c L L} 
        \textbf{Reconstruction} & \textbf{Data Fidelity}  &\textbf{MSE}        &\textbf{LPIPS}      &\textbf{Time on GPU (ms)}      & \textbf{\# Training Images}  \\ \hline
        ADMM (converged)    &    13.62           &      .0622       &   .5711   & 1,520  &  none      \\ 
        ADMM (bounded)      &    11.32           &      .1041       &   .6309   & 71    &  none                  \\ \hline
        \textbf{Le-ADMM}              &    13.70           &      .0618       &   .4434   & 71    &   100             \\ 
        \textbf{Le-ADMM*}             &    16.21          &    .0309          &  .327    & 200      &   100              \\ 
        \textbf{Le-ADMM-U}      &   22.14      &  \textbf{.0074}      &   \textbf{.1904}   & 75    &  23,000               \\ \hline
        U-Net               &    19           &   .0154             &    .2461  & 10     &  23,000    \\ 
    \end{tabular}
    \label{tab:summary}
\end{table}

\subsection{Test set results}

Table~\ref{tab:summary} summarizes the reconstruction performance and speed of our learned networks on the test set.  Here we can see that our fastest networks (Le-ADMM and Le-ADMM-U) are 20$\times$ faster than classic reconstruction algorithms (ADMM converged) and have similar or better average MSE and LPIPS scores.  Le-ADMM* is slightly slower due to its inclusion of a CNN on the uncropped image in each unrolled layer, however is still an order of magnitude faster than converged ADMM. As we move on the scale from classic to deep (Le-ADMM $\rightarrow$ Le-ADMM* $\rightarrow$ Le-ADMM-U), our networks have better MSE and LPIPS scores, but have worse data fidelity.

\begin{figure}[h]
\centering\includegraphics[width=5.25in]{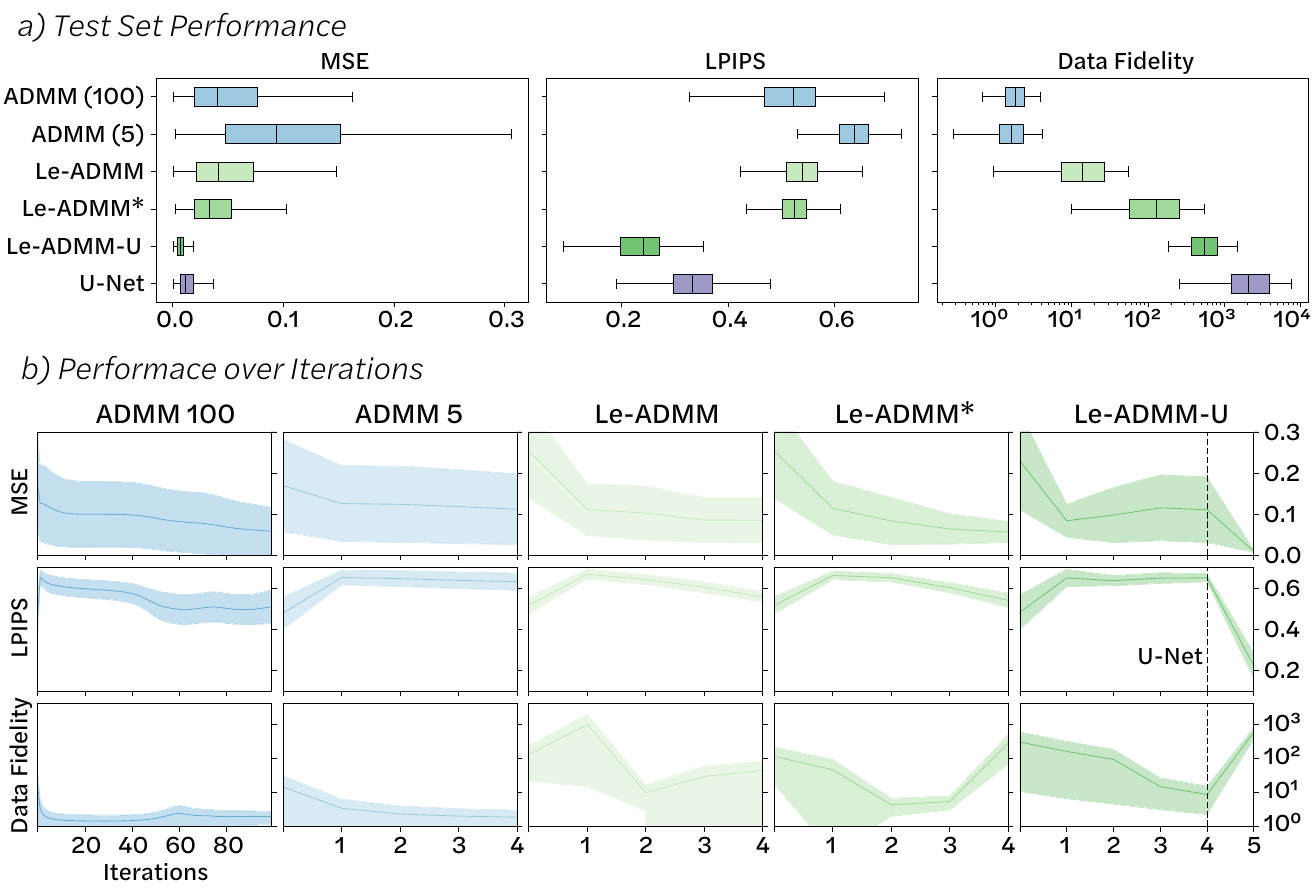}
\caption{Network Performance on Test Set. (a) Here we plot the MSE, LPIPS, and Data Fidelity values for all image pairs in our test set.  On average, our learned networks (green) are more similar to the ground truth lensed images (lower MSE and LPIPS) than 5 iterations of ADMM. Furthermore, our networks have comparable performance to ADMM (100), which takes 20$\times$ longer than Le-ADMM and Le-ADMM-U. However, the data fidelity term is higher for the learned methods, indicating that these reconstructions are less consistent with the image formation model. (b) Here we plot performance after each layer (or equivalently, each ADMM iteration) in our network, showing that MSE and LPIPS generally decrease throughout the layers.  The U-Net denoiser layer in Le-ADMM-U significantly decreases the LPIPS and MSE values, at the cost of data fidelity. }
\label{fig:lpips_mse}
\end{figure}

Figure~\ref{fig:screenresults} shows several sample images from our test set reconstructions.  Here we can see that our networks (Le-ADMM, Le-ADMM*, Le-ADMM-U) produce images that are of equal or better quality than converged ADMM.  We can see that bounded ADMM has streaky artifacts, but our learned networks do not. Le-ADMM-U has the best reconstruction performance overall and produces images that are visually similar to the ground truth images.  Overall, Le-ADMM-U has 3$\times$ better image quality than converged ADMM as measured by the LPIPS metric.  The U-Net does not perform as well as Le-ADMM-U, having inconsistent colors and missing higher frequencies.  This shows the utility in combining model-based and deep methods. 

Figure~\ref{fig:lpips_mse}(a), plots the distribution of MSE, LPIPS, and Data Fidelity scores for the test set.  We can see that Le-ADMM-U has the best LPIPS and MSE scores and outperforms converged ADMM, whereas Le-ADMM has similar LPIPS and MSE scores to converged ADMM with many fewer training pairs. Here we can clearly see the trend of data fidelity increasing as MSE and LPIPS decrease, showing that there is a trade-off between image quality and matching the imaging model.  We interpret this as our system model being imperfect, which prevents purely model-based algorithms from achieving the best image quality.  As we increase the number of learned parameters, we are able to correct artifacts introduced by model mismatch, producing more visually appealing images that better match the lensed camera. Figure~\ref{fig:lpips_mse}(b) analyzes what happens to the reconstruction throughout the layers of the learned network.  The MSE and LPIPS scores tend to decrease with iterations, while data fidelity increases.  For Le-ADMM-U, the U-Net greatly improves the LPIPS and MSE values, at the cost of data fidelity. 

\begin{figure}[t]
\centering\includegraphics[width=5.25in]{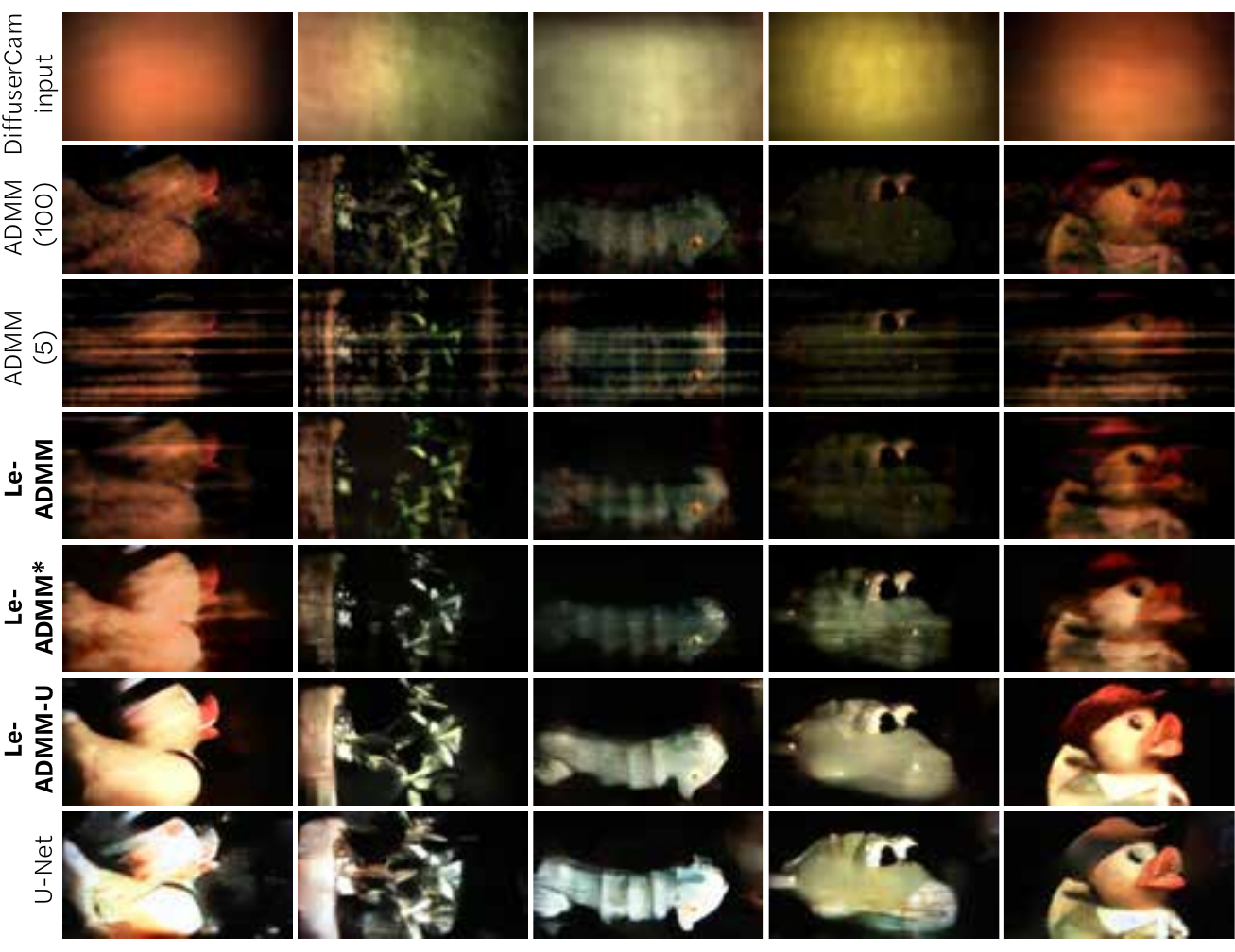}
\caption{Network performance on objects in the wild (toys and a plant) captured with our lensless camera.  We show the raw measurement (contrast stretched) on the top row, followed by converged ADMM, ADMM bounded to 5 iterations, our learned networks, and U-Net for comparison. Our learned networks have similar or better image quality as converged ADMM, and Le-ADMM-U has the best image quality.  For instance, Le-ADMM-U is able to capture the details in the sideways plant (second column from left) and the eye of the toy duck (right). The U-Net alone has good image quality, but is missing some colors and details (e.g. the first image is washed out and the nose of the alligator toy is miscolored). }
\label{fig:inthewild}
\end{figure}

\subsection{Generalization to images in the wild}
Next, we remove the computer monitor and capture DiffuserCam images of natural objects. Figure~\ref{fig:inthewild} shows some example reconstructions using our learned networks. Again, we see that our networks  produce images of similar or higher visual quality than converged ADMM.  In particular,  Le-ADMM-U again produces the most visually appealing images and has better image quality than converged ADMM. This shows that our learned networks are able to generalize beyond imaging a computer monitor to situations with dramatically different lighting conditions.

\section{Discussion}
Our work presents a preliminary analysis of using unrolled, model-based neural networks on a real experimental lensless imaging system.  We show that it is favorable to choose a network that combines classic and deep methods.  We can perform comparably to classic algorithms at a fraction of the speed using only a few learned parameters, but can greatly improve image quality when increasing the number of learned parameters. However, the number of learned parameters in the network could be varied depending on the application.  For instance, scientific imaging applications might choose to have fewer learned parameters to prevent overfitting to the training data. Meanwhile, photography applications may prefer a deeper method with more parameters, potentially producing more visually appealing images at the expense of possibly hallucinating details not present in the scene. 

The quality and resolution of our reconstructions is bounded by that of our training dataset, including any potential imperfections in the physical system.  For instance, any aberrations introduced by our lensed camera or beamsplitter will affect the learned reconstructions, since the lensed images are used as the ground truth when updating the network parameters.  However, in practice we correct for aberrations such as distortion before training; other effects (e.g. chromatic aberration, field curvature) are negligible at our reconstruction grid size.  Possible future work includes training on scenes with larger depth content to yield reconstructions with desirable defocus blurs, such as seen in a lensed camera.

\section{Conclusion}
We presented several unrolled, model-based neural networks for lensless imaging with a varying number of trainable parameters.  Our networks jointly incorporate the physics of the imaging model as well as learned parameters in order to use both the known physics and the power of deep learning.  We presented an experimental system with a prototype lensless camera that was used to rapidly acquire a dataset of aligned lensless and lensed images for training.  Each of our networks are able to produce similar or better image quality compared to standard algorithms, with the fastest offering a 20$\times$ improvement.  In addition, our deeper method, Le-ADMM-U has 3$\times$ better image quality than standard algorithms on the LPIPS perceptual similarity scale. Our learned network is fast enough for interactive previewing of the scene and also produces visually appealing images, addressing two of the big limitations of lensless imagers. Our work suggests that using such model-based neural networks could greatly improve imaging speed and quality for lensless imaging at the cost of a training step before camera operation.

\section*{Acknowledgments}
Kristina Monakhova and Kyrollos Yanny acknowledge support from the NSF Graduate Research Fellowship Program.  Grade Kuo is a National Defense Science and Engineering Graduate Fellow.  The authors thank Ben Mildenhall for helpful discussions.  

\section*{Funding}
This material is based upon work supported by the National Science Foundation Graduate Research Fellowship (DGE 1752814); STROBE: A National Science Foundation Science \& Technology Center (DMR 1548924); Gordon and Betty Moore Foundation's Data-Driven Discovery Initiative (GBMF4562)

\section*{Appendix A}
\subsection*{Network architecture}
We outline our U-Net network architecture (used for Le-ADMM-U as well as for the U-Net comparison) below in Table~\ref{tab:unet} for completeness.  This is based on the architecture specified in \cite{ronneberger2015u}. 
\renewcommand{\arraystretch}{1.0}
\begin{table}[h]
\caption{Network architecture for the U-Net used in Le-ADMM-U.  Here \textbf{k} represents the kernel size, \textbf{s} is the stride, \textbf{channels in/out} represents the number of input and output channels for the layer, and \textbf{input} is the input of the layer, with ',' representing concatenation.  Here the encoding steps, enc, consist of two convolutional layers, each of which consists of a 2D convolution, followed by a batch-norm and ReLu.  The decoding steps, dec, consists of three convolutional layers with the same architecture. Here, up$(\cdot)$ stands for bilinear upsampling.  conv1 consists of a convolutional layer, batch-norm, and ReLu, whereas conv2 consists only of a convolutional layer. }
\vspace{2mm}
\centering
\begin{tabular}{ c c c c c  } 
 \hline
\textbf{layer} & \textbf{k} & \textbf{s}& \textbf{channels in/out} & \textbf{input}  \\ 
\hline 
enc1 & 3 & 1& $3/24$ & input\\
pool1  &  2 & 2 &  $24/24$ & enc1\\
enc2 & 3 & 1& $24/64$ & pool1 \\
pool2   & 2  & 2 &  $64/64$ & enc2\\
enc3 & 3 & 1& $64/128$ & pool2\\
pool3   & 2  & 2 &  $128/128$ & enc3\\
enc4 & 3 & 1& $128/256$ & pool3\\
pool4   & 2  & 2 &  $256/256$ & enc4\\
enc5 & 3 & 1& $256/512$ &pool4\\
pool5   & 2  & 2 &  $512/512$ & enc5\\
conv1 & 3 &1 & $512/512$ &pool5\\ \hline
dec5 & 3 & 1& $512/256$ & up(conv1), enc5\\
dec4 & 3 & 1& $256/128$ & up(dec5) enc4\\
dec3 & 3 & 1& $128/64$ & up(dec4), enc3\\
dec2 & 3 & 1& $64/24$ & up(dec3), enc2\\
dec1& 3 & 1& $24/24$ & up(dec2), enc1\\
conv2 & 1 & 1 & $24/3$ & dec1 \\
 \hline
\end{tabular}
\label{tab:unet}
\end{table}

Next, we outline our smaller U-Net that is used for Le-ADMM*.  The network architecture is described as follows: 
\renewcommand{\arraystretch}{1.0}
\begin{table}[h]
\centering
\caption{Network architecture for smaller U-Net that is used in Le-ADMM*.  The encoding and decoding steps are the same as described in Table \ref{tab:unet}.   Finally, we include a skip connection, adding the input of the network to the output. }
\vspace{2mm}
\begin{tabular}{ c c c c c  } 
 \hline
\textbf{layer} & \textbf{k} & \textbf{s}& \textbf{channels in/out} & \textbf{input}  \\ 
\hline 
enc1 & 3 & 1& $3/24$ & input\\
pool1  &  2 & 2 &  $24/24$ & enc1\\
conv1 & 3 &1 & $24/24$ &pool1\\ \hline
dec1& 3 & 1& $24/24$ & up(conv1), enc1\\
conv2 & 1 & 1 & $24/3$ & dec1 \\
 \hline
\end{tabular}
\label{tab:unet_small}
\end{table}

\subsection*{Effect of training size}
\begin{figure}[h]
\centering\includegraphics[width=3in]{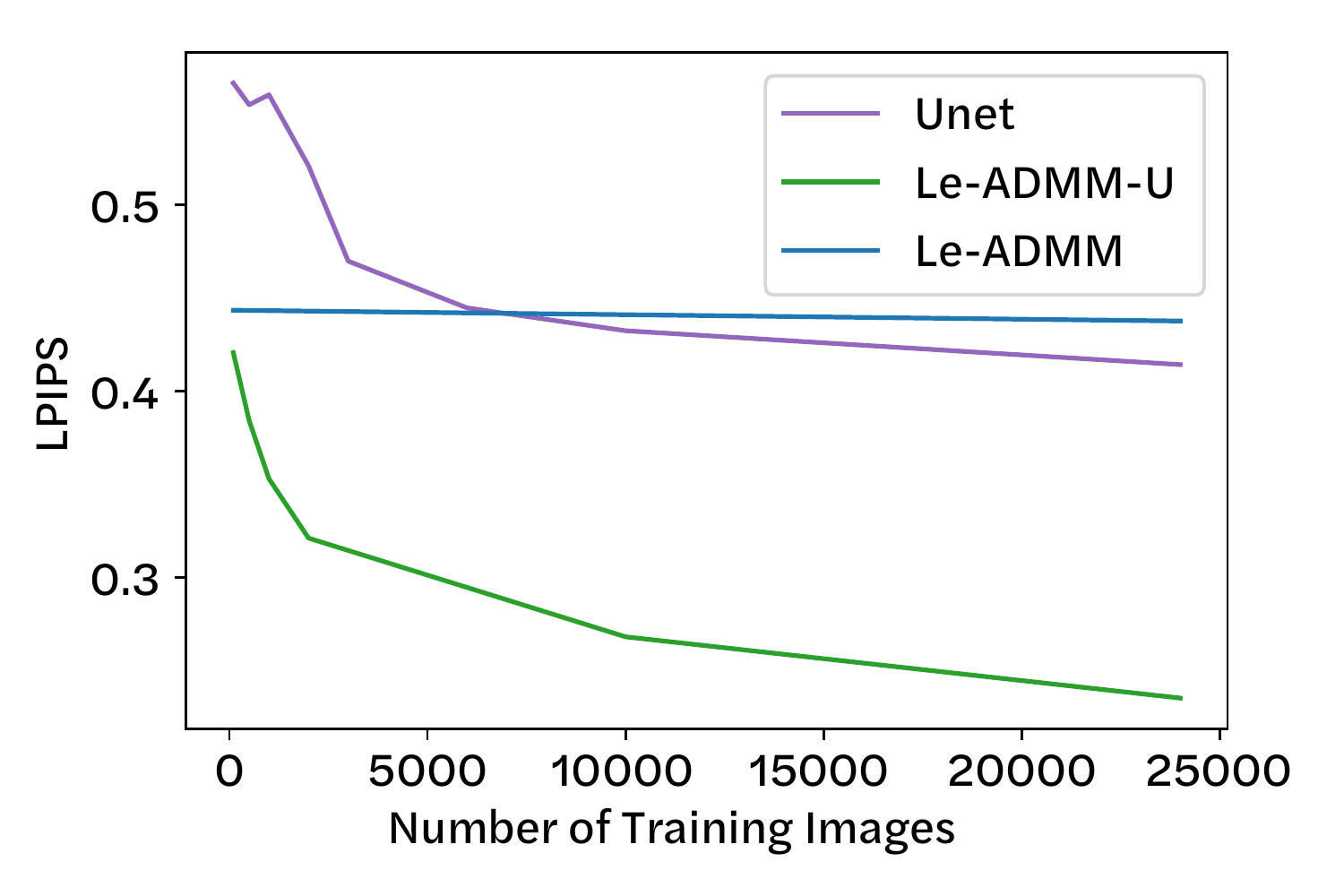}
\caption{Effect of Training Size.  Here we vary the number of images in the training set and plot the LPIPS score after 5 epochs.  Here we see that Le-ADMM-U performs better and converges faster than a U-Net alone.  Le-ADMM does not improve as the number of training images increases, since it has so few parameters.}
\label{fig:trainingsize}
\end{figure}

In Fig.~\ref{fig:trainingsize} we study the effect of the number of training images on the network performance.  We show that our model-based network, Le-ADMM-U, is able to perform much better than the deep method (U-Net) with fewer training images because it incorporates knowledge of the imaging system into the network. 

\bibliography{bib}

\end{document}